**Final report (excerpt), Strategic Partnership Grants, NSERC, STPGP 447326 – 13**

Research on

# Fiber-shaped lithium-ion batteries with metallic electrodes

described in this final report excerpt was conducted as part of the abovementioned Strategic Partnership Grant.

All the research activities detailed in this final report excerpt were conducted exclusively in the laboratory of Dr. Skorobogatiy, Polytechnique Montréal.

All the data presented in this report was obtained and interpreted exclusively by the research associate Dr. H. Qu with the help of the graduate student X. Lu under immediate supervision of Dr. M. Skorobogatiy.

The following text of the report was written by Dr. H. Qu with the help of X. Lu and under immediate supervision of M. Skorobogatiy

**Acknowledgements:** While our research group worked largely independently from the group of co-PI O. Semenichin, and while the overall battery structure and fabrication techniques were originally conceived at Ecole Polytechnique using our prior experiences in solid electrolite Li-ion batteries (see ref. [*]), we would like to acknowledge that our co-PI provided us occasionally with his insights into practical aspects of chemical processing, electrochemical characterization and material's processing. The choice of Al as one of the possible battery electrode materials was also originally suggested by the co-PI. Morover, co-PI has visited our group several times to observe the work conducted in our labs, while Dr. H. Qu has visited our co-PI reserch lab once as part of the collaborative exchange of ideas and practices.

* - Y. Liu, S. Gorgutsa, C. Santato, and M. Skorobogatiy "Flexible, Solid Electrolyte-Based Lithium Battery Composed of LiFePO4 Cathode and Li4Ti5O12 Anode for Applications in Smart Textiles," Journal of The Electrochemical Society, vol. 159 (4), pp. A349-A356 (2012)

**Final report (excerpt), Strategic Partnership Grants, NSERC, STPGP 447326 – 13**

**Summary of the conducted research**

We have demonstrated a fiber lithium ion battery (LIB) fabricated by co-twisting a $LiFePO_4$ composite-coated copper wire (cathode) together with an aluminum wire (anode). An all-solid $LiPF_6$ composite layer functioning both as the electrolyte and battery separator is deposited onto the two electrode wires before twisting. To characterize the electrochemical properties of the battery, charge-discharge tests with different C-rates are performed. The fiber LIB has an open-circuit voltage of ~3.4 V, and the typical specific capacity is found to be ~87 $mAhg^{-1}$ at 0.5 C charge-discharge rate. Besides, the proposed battery has a Coulombic efficiency of more than 82% throughout all the charge-discharge tests. We also find that bending of the fiber battery has insignificant influence on the battery electrochemical properties

**Fabrication of fiber lithium ion batteries**

To fabricate the LIB cathode wire, a LFP-PVDF composite solution was first prepared by dissolving LFP and PVDF into 1-Methyl-2-pyrrolidinone (NMP) solvent. Carbon black was added into the cathode solution to increase the conductivity. Then, the LFP composite layer was deposited onto a copper wire using dip-and-dry method in a $N_2$-filled glove box. The obtained cathode wire was then deposited with an electrolyte layer also using dip-and-dry method (Fig. 1(b)). The electrolyte solution was prepared by dissolving $LiPF_6$ and PEO into acetonitrile solvent. Particularly, $TiO_2$ was also added into the electrolyte solution in order to lower the polymer crystallinity and improve the electrolyte ionic conductivity [17]. The LIB anode was simply an aluminum wire coated with an electrolyte layer using dip-and-dry method (Fig. 1(a)). We then twisted two electrode wires using a home-made jig fabricated using a Makerbot 3D printer. In order to enhance the bonding of the electrolyte polymer (PEO) matrix on the two electrode wires, we then wetted the battery with several drops of propylene carbonate. Finally, the battery was encapsulated with a heat-shrinkable tube, and was heated at 120 ℃ for 60 seconds. Note that the dip-and-dry process and battery assembly were all performed inside the N$_2$-filled glove box.

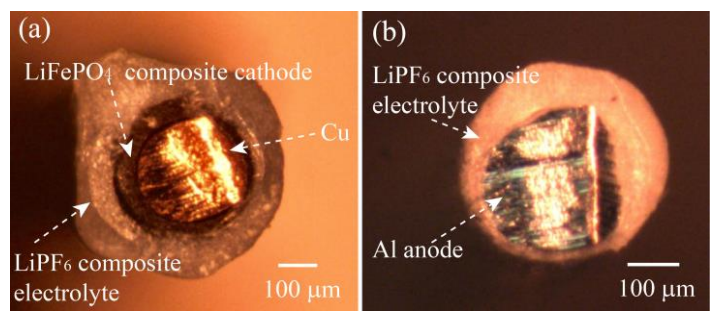

Figure 1. (a) Cross section of the LIB cathode wire; (b) Cross section of the LIB anode wire.

**Characterization of electrochemical properties of lithium ion batteries**

The electrochemical properties of a fiber LIB were investigated using cyclic charge-discharge analysis with different charge-discharge rates (from 0.5 C to 8 C) as shown in Fig. 2 and 3. Experimentally, a 15 cm-long fiber battery was fabricated. Then, the fiber battery was cut into two samples (Sample 1 and Sample 2) with equal length. Both extremities of each sample were sealed with epoxy. The charge-discharge measurements for Sample 1 were carried out inside the N$_2$-filled glove box using an Ivium Electrochemical Workstation (Fig. 2(a, b)), while the corresponding tests for Sample 2 were performed outside the glove box (in a regular laboratory environment). The open-circuit voltage of both samples was found to be ~ 3.4 V. As shown in Fig. 2(c, d), Sample 1 had specific capacities of 87.1, 57.7, 37.2, 29.3, 23.0 mAhg$^{-1}$ at 0.5, 1, 2, 4, and 8 C charge-discharge rate, respectively. We note that the battery capacities decrease rapidly with increased current rates (Peukert constant of ~2). This is because the practical ionic current density of the electrolyte and electrodes, including the rate of ion transfer across the electrode/electrolyte interface, is generally much smaller than the electronic current density of the external electronic circuit. Therefore, at a high charge/discharge current rate, the ionic motion within an electrode and/or across an electrode/electrolyte interface is too slow for the charge distribution to reach equilibrium, thus leading to a decreased capacity. As the charge-discharge current rate decreased, the capacity loss was also recovered. Besides, we like to mention that this LIB still has the voltage plateau that remained flat even at a current rate as high as 8 C, indicating good charge-discharge performance[18]. After a series of tests with different C-rates, the discharge capacity of the LIB was ~ 56.6 mAhg$^{-1}$, when the rate turned back to 1 C. This indicates that the structure of each components of the full battery remained intact after subjecting to high current densities. Finally, 25 cycles of charge-discharge tests at 1 C rate were performed. The specific capacity of the LIB was consistent in the 25-cycle tests. The coulombic efficiencies of the battery were greater than 82% during the whole charge-discharge cycles (Fig. 2(d)), while mostly staying above ~89%.

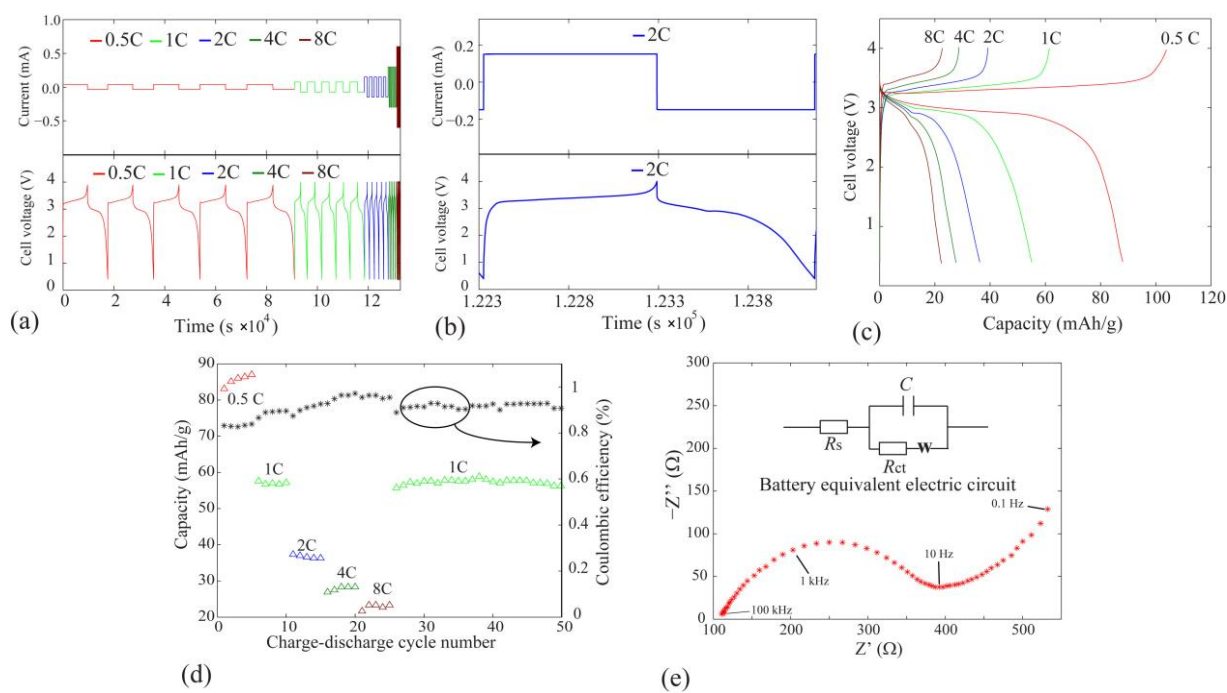

Figure 2. Electrochemical performance of Sample 1. (a) Cell voltages and currents in the first 25 cyclic charge-discharge tests with different C-rates. (b) A zoomed view of the cell voltage and current in the charge-discharge test with a 2-C rate. (c) Voltages of Sample 1 in different charge-discharge cycles as a function of the battery capacity. (d) Specific capacity and coulombic efficiency of Sample 1 measured in 50 charge-discharge cycles with different C rates. (e) EIS spectrum of Sample 1 (The inset is the equivalent electric circuit of the battery).

The electrochemical impedance spectroscopy (EIS) measurement of Sample 1 was then performed. The Nyquist plot of Sample 1 (shown in Fig. 2(e)) is composed of a depressed semicircle in the high-to-medium frequency region followed with a slope in the low frequency region. According to the order of descending frequency, the EIS spectrum could be divided into three distinct regions. This is a classic shape of the EIS that can be fitted using an effective electric model shown in the insert of Fig. 3(a). There, $R$ denotes various ohmic resistances, while $W$ denotes Warburg impedance. The first intercept on the real axis in Fig. 2(e) gives the equivalent series resistance, $R_s$, which is a bulk electrolyte resistance. The second intercept gives a sum of the electrolyte resistance, $R_s$ and the charge transfer resistance, $R_{ct}$, which is the electrode-electrolyte interfacial resistance. From Fig. 2(e), $R_s$ and $R_{ct}$ of Sample 1 are ~115 Ω and ~276 Ω, respectively.

Fig. 3 (a-d) shows the results of the charge-discharge tests of Sample 2 which was measured outside of the glove box. We note that the electrochemical performance of Sample 2 is well compared to that of Sample 1. The specific capacities of Sample 2 are 85.6, 53.0, 34.3, 22.5, 9.2 mAhg$^{-1}$ at 0.5, 1, 2, 4, and 8 C, respectively. The coulombic efficiency of Sample 2 is greater than 84% for all of the charge-discharge cycles, while most staying over 88%. Besides, $R_s$ and $R_{ct}$ of Sample 2 are ~155 Ω and ~345 Ω, respectively (Fig. 3(e)).

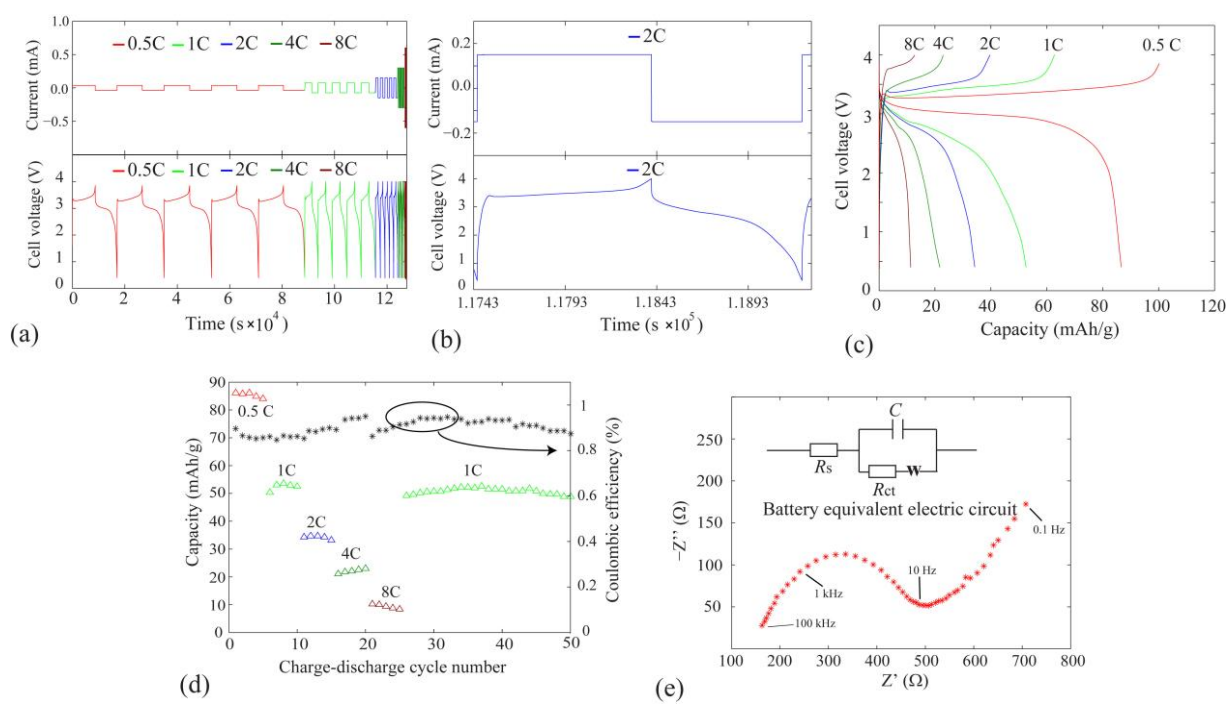

Figure 3. Electrochemical performance of Sample 2. (a) Cell voltages and currents in the first 25 cyclic charge-

discharge tests with different C-rates. (b) A zoomed view of the cell voltage and current in the charge-discharge test with a 2-C rate. (c) Voltages of Sample 2 in different charge-discharge cycles as a function of the battery capacity. (d) Specific capacity and coulombic efficiency of Sample 2 measured in 50 charge-discharge cycles with different C-rates. (e) EIS spectrum of Sample 2.

To study how bending of a fiber battery would affect its electrochemical performance, we fabricated another 15 cm-long fiber battery, and performed the cyclic charge-discharge tests (with a 2 C rate) of this fiber battery as a function of different bending curvatures. Experimentally, one end of this fiber battery was fixed, while the other end was displaced by 0, 1, 2, and 3 cm, respectively, using a micro-positioning stage (Fig. 4(a, b)). At each displacement position, 5 cycles of charge-discharge tests were carried out. Finally, the moving end was pulled back to its original position for another charge-discharge measurement. From the results shown in Fig. 4(c, d), we found that bending of the fiber battery has insignificant influence on the battery performance. The capacity of the fiber battery decreased by ~3.8% after 25 cycle bending tests.

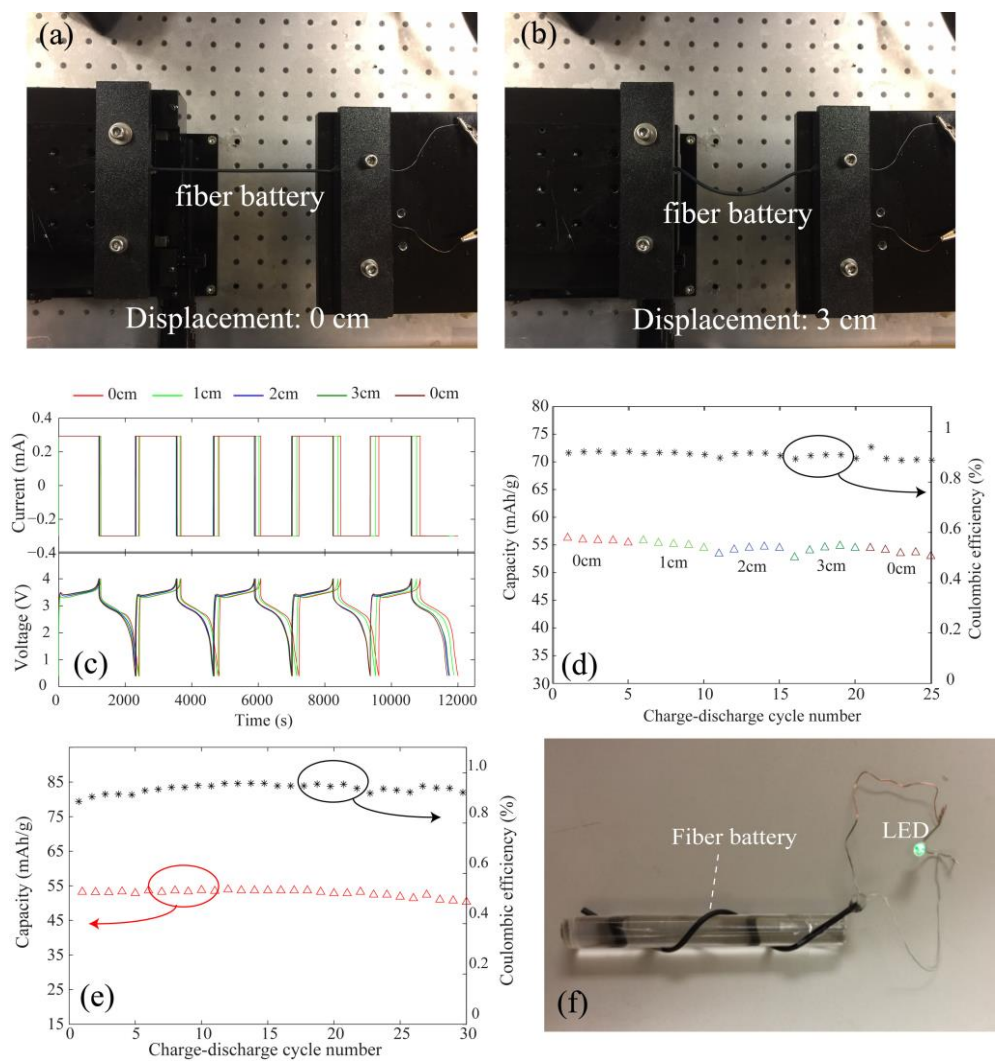

Figure 4. Electrochemical performance of a 15-cm long fiber battery as a function of bending. (a) and (b) show the setup for bending the fiber battery. (c) Charge-discharge tests of the fiber battery at a 2 C rate when the moving end of the battery was displaced to different positions. (d) Specific capacity and coulombic efficiency of the fiber battery in the bending tests. (e) Specific capacity and coulombic efficiency of the fiber battery that underwent repeated bend-release motions. Particularly, the moving end of the fiber battery was displaced by 2 cm and then pulled back to its original position once every 4 s. Totally, more than 16000 bend-release actions of the battery were implemented. (f) A 15 cm-long fiber LIB wrapped around an acrylic rod (diameter: ~1.27 cm) was used to light up a green LED (working parameters: 20 mA @ 3.3 V).

We also performed a durability test of the same fiber battery regarding to bending. In the test, the moving end of the fiber battery was displaced by 2 cm and then pulled back once every 4 s, while the battery was charged and discharged at a 2 C rate. Totally, ~ 16000 bend-release actions were implemented during 30 cycles of charge-discharge tests. The specific capacity and Coulombic efficiency of the fiber battery virtually remain consistent throughout the whole test as shown in Fig. 4(e).

In Fig. 4(f), we demonstrate the flexibility of the fiber LIB. A 15 cm-long fiber LIB wound around an acrylic rod with a diameter of 1.27 cm was used to light up a green LED (operation parameters: 20mA@3.3V, Thorlabs).

**Conclusions**

Within this strategic project, we reported an all-solid, fiber LIB that has a $LiFePO_4$ composite cathode wire twisted with an aluminum anode wire. The cathode wire is fabricated by depositing a $LiFePO_4$ composite layer and a $LiPF_6$-composite layer on a copper-wire substrate using a dip-and-dry method. The anode wire is fabricated by depositing the $LiPF_6$-composite layer on an aluminum wire also using the dip-and-dry method. Then, the two electrode wires are twisted together using a home-made jig to constitute a full LIB. The proposed LIB shows an open-circuit voltage of ~3.3 V. The experimental results of the charge-discharge tests suggest that the battery has a specific capacity of ~87 $mAhg^{-1}$ at 0.5 C rate. Moreover, the battery could retain well its capacity after intensive cyclic charge-discharge operation. During all the battery operation, the coulombic efficiency of the battery remains above ~82%. We also find that the electrochemical properties of the fiber battery is virtually independent of bending actions. The fabrication of the battery is extremely simple and economic, and all of the battery component materials are inexpensive and commercially available. Compared to the majority of currently existing fiber- (cable-) shaped LIBs that utilize electrolytes based on liquid organic solutions, the proposed LIB is much more advantageous for wearable applications thanks to its all-solid structure.

**Experimental Details**

### (1) Fabrication of anode wire

0.125 g $LiPF_6$ (powder, Sigma-Aldrich) and 0.08 g $TiO_2$ (nanopowder, Sigma-Aldrich) was dispersed into 12.5 ml acetonitrile (99.9%, Sigma-Aldrich) solvent for 3 hour using a magnetic stirrer at room temperature (22 ℃). Then, 0.665 g polyethylene oxide (powder, Mw ~400,000, Sigma-Aldrich) was added into the solution which was then stirred for 12 hours at room temperature (22 ℃). The as-prepared solution will be casted onto an aluminum anode wire to form an electrolyte layer which also function as the battery separator. All these processes were carried out in the $N_2$-filled glove box. An aluminum wire (diameter: 0.33 mm, 99% purity, Mcmaster Carr) was manually polished using sanding sheets (Grit 1000, Mcmaster Carr), and then rinsed with regular detergent, 1 wt% KOH solution, and isopropanol for 20 minutes each in an ultrasonic bath. After rinsing, the aluminum wire was blown to dry with $N_2$ flow, and then was stored in a $N_2$-filled glove box. The LIB anode wire was then fabricated by depositing a $LiPF_6$ composite layer onto the aluminum wire via a dip-and-dry process in the $N_2$-filled glove box.

### (2) Fabrication of cathode wire

A polyvinylidene fluoride (PVDF) solution was first prepared by dispersing 1g PVDF (powder, Mw. ~534,000, Sigma-Aldrich) into 10 ml 1-Methyl-2-pyrrolidinone (NMP) solvent (99.5%, Sigma-Aldrich) using a magnetic stirrer for 2 hours at room temperature (22 ℃). 0.425 g LFP (Phostech Lithium Inc.) was manually ground for 10 minutes and then mixed with 0.025 g carbon black (Super P, Alfa Aesar). The mixture was then dispersed into 2 ml as-prepared PVDF solution using the magnetic stirrer for 6 hours at room temperature (22 ℃). A copper wire (diameter: 0.33 mm, 99% purity, Mcmaster Carr) was first manually polished using sanding sheets (Grit 1000, Mcmaster Carr), and then rinsed with regular detergent, 1 wt% HCl solution, and isopropanol for 20 minutes each in an ultrasonic bath. After rinsing, the copper wire was blown to dry with $N_2$ flow, and then was stored in a $N_2$-filled glove box. The cathode wire was fabricated by depositing a LFP composite layer onto the copper wire via a dip-and-dry process in the $N_2$-filled glove box. Finally, an electrolyte layer was also deposited onto the cathode wire using dip-and-dry method.

(3) Fabrication of LIB

The cathode and anode wires were co-twisted using a home-made jig fabricated with a Makerbot 3D printer. The twisted LIB was then wetted by several drops of propylene carbonate. Finally, the LIB was encapsulated with a heat-shrinkable tube and was heated at 120 ℃ for 60 seconds. Both ends of the LIB were then sealed with epoxy.

(4) Battery conditioning

Before performing the charge-discharge tests of each battery, we need to carry out a battery conditioning process which would allow the formation of the nanostructures on the electrode that would withstand (or partially withstand) the volume change caused by the lithiation-delithiation process during the battery operation [19]. In the conditioning process, we generally used a 0.5 C current (calculated from the theoretical capacity of $LiFePO_4$) to charge-discharge the battery with different charging periods. In particular, we started with a charging period of 1 min, and charge-discharged the battery for 4 cycles. Then, we gradually increased the charging period from 1 min to 30 min with a step of 5 min, and for each charging period we run 4 charge-discharge cycles. During the conditioning process, we could also see that the coulombic efficiency of the battery increased gradually from ~60% to higher than 80%.